\documentclass[preprint,showpacs,preprintnumbers,amsmath,amssymb]{revtex4}

\usepackage{color}
\usepackage{graphicx}

\usepackage{dcolumn}
\usepackage{bm}
\usepackage{lscape}
\usepackage{ulem}
\usepackage{mathrsfs}
\usepackage{braket}

\usepackage{url}

\usepackage{txfonts}

\begin{document}


\title{Effect of channel coupling on the elastic scattering of lithium isotopes}

\author{T.~Furumoto}
\email{furumoto-takenori-py@ynu.ac.jp}
\affiliation{Graduate School of Education, Yokohama National University, Yokohama 240-8501, Japan}

\author{T.~Suhara}%
\email{suhara@matsue-ct.ac.jp}
\affiliation{Matsue College of Technology, Matsue, Shimane 690-8518, Japan}

\author{N.~Itagaki}%
\email{itagaki@yukawa.kyoto-u.ac.jp}
\affiliation{Yukawa Institute for Theoretical Physics, Kyoto University, Kyoto 606-8502, Japan}

\date{\today}

\begin{abstract}
Herein, we investigated the channel coupling (CC) effect on the elastic scatterings of lithium (Li) isotopes ($A =$ 6--9) for the $^{12}$C and $^{28}$Si targets at $E/A =$ 50--60 MeV.
The wave functions of the Li isotopes were obtained using the stochastic multi-configuration mixing (SMCM) method based on the microscopic-cluster model.
The proton radii of the $^{7}$Li, $^{8}$Li, and $^{9}$Li nuclei became smaller as the number of valence neutrons increased.
The valence neutrons in the $^{8}$Li and $^{9}$Li nuclei exhibited a glue-like behavior, thereby attracting the $\alpha$ and $t$ clusters.
Based on the transition densities derived from these microscopic wave functions, the elastic-scattering cross section was calculated using a microscopic coupled-channel (MCC) method with a complex $G$-matrix interaction.
The existing experimental data for the elastic scatterings of the Li isotopes and $^{10}$Be nuclei were well reproduced.
The Li isotope elastic cross sections were demonstrated for the $^{12}$C and $^{28}$Si targets at $E/A$ =53 MeV.
The glue-like effect of the valence neutrons on the Li isotope was clearly demonstrated by the CC effect on elastic scattering.
Finally, we realize that the valence neutrons stabilized the bindings of the core parts and the CC effect related to core excitation was indeed reduced.
\end{abstract}


\maketitle

\section{Introduction}

Recently, study of unstable nuclei has significantly advanced both from theoretical and experimental viewpoints.
Many new phenomena attributed to the increase in the degrees of freedom with the addition of valence protons or neutrons have been discussed in various unstable nuclei.
For instance, halo structure is a characteristic feature of valence neutrons that are weakly bound around core nuclei~\cite{TAN85, TAN85-2, MIN92}.
Additionally, the change or inversion of a single-particle structure~\cite{DOO09, DOO10, SAN15} and  Pygmy giant resonance~\cite{ADR05, SAV06, WIE09} have been observed as exotic structures of unstable nuclei.

In light neutron-rich nuclei, it has been shown that valence neutrons exhibit a glue-like behavior; they strengthen the bindings of the clusters.
The $^{8}$Be ($\alpha + \alpha$) nucleus is well known as an unbound system; however, the $^{9}$Be ($\alpha + \alpha + n$) nucleus is bound by the addition of a neutron.
The valence neutrons play an important role in stabilizing the $^{10}$Be and $^{12}$Be nuclei~\cite{KIM16}.
Additionally, the valence neutrons in molecular orbits contribute to the binding.
In addition, various molecular and atomic orbit configurations appear as the excited states of these nuclei~\cite{ITA00, ITO08}.
Furthermore, in neutron-rich C isotopes, the valence neutrons play an important glue-like role in the stabilization of the three-$\alpha$-cluster states, including the linear-chain configurations~\cite{ITA04, SUH10}.

In terms of the observable appearance of this glue-like effect, in nuclear reactions, it is difficult to compare the $^{8}$Be and $^{9}$Be nuclei because they are unbound and bound nuclei, respectively.
In contrast, Li isotopes are good candidates for observing the glue-like effect.
First, the loosely bound nature of the $^{6}$Li and $^{7}$Li nuclei has been thoroughly investigated; they are well known as $\alpha$+$d$ ($\alpha$+$p$+$n$) and $\alpha$+$t$ cluster systems, respectively.
The breakup effects of the $^{6}$Li and $^{7}$Li nuclei into $\alpha$+$d$ ($\alpha$+$p$+$n$) and $\alpha$+$t$ systems, respectively, such as the channel coupling (CC) effect on elastic scattering, have been well investigated~\cite{SAK86, SAK87, WAT15}.
The scattering reaction data provides detailed information about the nuclear structure and nuclear reaction mechanisms.
For instance, the CC effect on the breakup reaction of the $^{6}$He, $^{6}$Li, and $^{7}$Li nuclei~\cite{SAK86, SAK87, MAT04, WAT15}, the contribution of the core excitation to the elastic and inelastic scatterings~\cite{HOR10, MOR12-2}, the important role of the multi-step reaction in inelastic scattering~\cite{TAK09}, and the decoupling of the deformation of the proton and neutron densities~\cite{TAK05} have been investigated.

We also analyze the elastic scattering of the Li isotopes to investigate the role of its valence neutrons.
If valence neutrons are added to the $^{7}$Li nucleus, they are anticipated to have an impact on the CC effect.
As mentioned above, Li isotopes are good candidates for investigating the glue-like behavior of valence neutrons in nuclear reactions.
As for the framework, we combine microscopic nuclear structure and reaction calculations.
For the structure part, we first employ the stochastic multi-configuration mixing (SMCM) method with the cluster structure to obtain the wave functions of the Li isotopes.
The SMCM method has been known to well describe the structure of light nuclei, not only for the ground state but also for the excited state~\cite{ICH11,MUT11}.
Next, using the transition densities based on the results of the SMCM method, the microscopic coupled-channel (MCC) method is employed to describe Li-isotope scattering from the viewpoint of the nuclear reaction in a manner same as that considered in previous research~\cite{FUR13-2}.
Herein, we investigate the glue-like behavior of the valence neutrons in this elastic scattering.

In this study, we first introduce the framework of the SMCM method for the calculating the densities of Li isotopes ($^{6}$Li, $^{7}$Li, $^{8}$Li, and $^{9}$Li).
Next, MCC calculation with a complex $G$-matrix interaction is introduced to calculate the elastic cross section.
The calculated excitation energy and the radius of the ground state are compared with the experimental data.
The existing experimental data for light heavy-ion elastic and quasi-elastic cross sections are analyzed in the MCC calculation based on the density calculated using the SMCM method.
The proposed model is employed to estimate the CC effect of the Li isotopes on elastic scattering.
Finally, the glue-like behavior of the valence neutrons in the $^{8}$Li and $^{9}$Li nuclei is verified in terms of nuclear reactions.

\vspace{3mm}
\section{Formalism}
\subsection{Microscopic cluster model}
We first introduced the SMCM method based on the microscopic cluster model~\cite{ICH11,FUR13-2} for calculating the densities of $^{6}$Li, $^{7}$Li, $^{8}$Li, and $^{9}$Li. 
Next, we introduced the basis state $\{ \Psi_{i}^{J^{\pi} MK} \}$ and diagonalized the Hamiltonian for each nucleus.
The total wave function $\Phi^{J^{\pi} M}$ can be expressed as follows:
\begin{equation}
\Phi^{J^{\pi} M} = \sum_{K} \sum_{i} c_{i, K} \Psi_{i}^{J^{\pi} MK}.
\label{mcmwf}
\end{equation}
The eigenstates of the Hamiltonian were obtained by diagonalizing the Hamiltonian matrix.
Additionally, the coefficients $\{c_{i, K} \}$ for the linear combination of Slater determinants were obtained.

Concretely, we prepared various $\alpha + p + n$ configurations for the basis states to describe the $^{6}$Li nucleus as follows:
\begin{eqnarray}
\Psi_{i}^{J^{\pi} MK} = &&  P^{\pi} P^{JMK} {\cal A} \nonumber \\
&& \big[ \phi_{\alpha} (\bm{r}_{1} \bm{r}_{2} \bm{r}_{3} \bm{r}_{4}, \bm{R}_{1}) \phi_{p} (\bm{r}_{5}, \bm{R}_{2}) \phi_{n} (\bm{r}_{6}, \bm{R}_{3}) \big]_{i},
\end{eqnarray}
where $\cal A$ is the antisymmetrizer and $\phi_{\alpha}$, $\phi_{p}$, and $\phi_{n}$ are the wave functions of $\alpha$, proton, and neutron, respectively.
Additionally, $\{ \bm{r}_{i} \}$ represents the spatial coordinates of the nucleons and each nucleon is described as a locally shifted Gaussian centered at $\bm{R}$ $\{\exp[-\nu(\bm{r}_i - \bm{R})^2]\}$ with a size parameter of $\nu = 1/2b^2$, where $b=$ 1.46 fm.
Here, the positions of the Gaussian-centered parameter $\bm{R}$ are randomly generated.
For the $^{7}$Li nucleus, the basis states were obtained using various $\alpha + p + n + n$ configurations, which are described as follows:
\begin{eqnarray}
\Psi_{i}^{J^{\pi} MK} = &&  P^{\pi} P^{JMK} {\cal A} \nonumber \\
&& \big[ \phi_{\alpha} (\bm{r}_{1} \bm{r}_{2} \bm{r}_{3} \bm{r}_{4}, \bm{R}_{1}) \phi_{p} (\bm{r}_{5}, \bm{R}_{2}) \nonumber \\
&& \ \phi_{n^{(1)}} (\bm{r}_{6}, \bm{R}_{3}) \phi_{n^{(2)}} (\bm{r}_{7}, \bm{R}_{4}) \big]_{i}.
\end{eqnarray}
To achieve a rapid energy convergence, we introduced two types of configurations for the basis states of the $^{8}$Li nucleus:,
\begin{eqnarray}
\Psi_{i}^{J^{\pi} MK} = &&  P^{\pi} P^{JMK} {\cal A} \nonumber \\
&& \big[ \phi_{\alpha} (\bm{r}_{1} \bm{r}_{2} \bm{r}_{3} \bm{r}_{4}, \bm{R}_{1}) \phi_{t} (\bm{r}_{5} \bm{r}_{6} \bm{r}_{7}, \bm{R}_{2}) \nonumber \\
&& \phi_{n} (\bm{r}_{8}, \bm{R}_{3}) \big]_{i},
\end{eqnarray}
where $\phi_{t}$ is the wave function of triton, and
\begin{eqnarray}
\Psi_{i}^{J^{\pi} MK} = &&  P^{\pi} P^{JMK} {\cal A} \nonumber \\
&& \big[ \phi_{\alpha} (\bm{r}_{1} \bm{r}_{2} \bm{r}_{3} \bm{r}_{4}, \bm{R}_{1}) \phi_{p} (\bm{r}_{5}, \bm{R}_{2}) \nonumber \\
&& \ \phi_{n^{(1)}} (\bm{r}_{6}, \bm{R}_{3}) \phi_{n^{(2)}} (\bm{r}_{7}, \bm{R}_{4}) \phi_{n^{(3)}} (\bm{r}_{8}, \bm{R}_{5}) \big]_{i}.
\end{eqnarray}
We mixed $\alpha + p + n + n + n$ and $\alpha + t + n$ configurations, where the formation of triton was assumed;
the contribution of the $\alpha + p + n + n + n$ basis states played only a minor role in the binding energy, radius, and transition strength.
For the $^{9}$Li nucleus, the formation of triton was assumed and the basis states were described by various $\alpha + t + n + n$ configurations~\cite{FUR13-2}, which are expressed as follows:
\begin{eqnarray}
\Psi_{i}^{J^{\pi} MK} = &&  P^{\pi} P^{JMK} {\cal A} \nonumber \\
&& \big[ \phi_{\alpha} (\bm{r}_{1} \bm{r}_{2} \bm{r}_{3} \bm{r}_{4}, \bm{R}_{1}) \phi_{t} (\bm{r}_{5} \bm{r}_{6} \bm{r}_{7}, \bm{R}_{2}) \nonumber \\
&& \ \phi_{n^{(1)}} (\bm{r}_{8}, \bm{R}_{3}) \phi_{n^{(2)}} (\bm{r}_{9}, \bm{R}_{4}) \big]_{i}.
\end{eqnarray}
The $\alpha$ cluster comprised four nucleons: spin-up proton, spin-down proton, spin-up neutron, and spin-down neutron.
These shared a common Gaussian-centered parameter, $\bm{R}_{1}$.
However, for simplicity, the spin and isospin of each nucleon were not explicitly described in this formula.
The triton comprised three nucleons (proton, spin-up neutron, and spin-down neutron) with the same Gaussian-centered parameters.
The projection onto an eigenstate of parity and angular momentum operators (projection operators $P^{\pi}$ and $P^{JMK}$) was performed numerically. 
The number of mesh points for the Euler angle integral was $16^3$, i.e., $4,096$.
The value of $M$ represents the $z$ component of the angular momentum in the laboratory frame, and the energy does not depend on $M$; however, it does depend on $K$, which is the $z$ component of the angular momentum in the body-fixed frame.

The Hamiltonian operator $H$ has the following form:
\begin{equation}
\displaystyle H=\sum_{i= 1}^{A} t_{i} - T_{\rm{c.m.}}+\sum_{i> j}^{A} v_{ij},
\end{equation}
where the two-body interaction $v_{ij}$ includes the central, spin-orbit, and Coulomb parts.
With regard to the nucleon-nucleon ($N$-$N$) interaction, we used the Volkov No.2 effective potential for the central part~\cite{VOL65}, which is expressed as follows:
\begin{eqnarray}
V( r)=&&( W-MP^{\sigma}P^{\tau}+ BP^{\sigma}-HP^{\tau})\nonumber \\&&
\times
( V_{1}\exp(-r^{2}/ c_{1}^{2})+ V_{2}\exp(-r^{2}/ c_{2}^{2})),
\end{eqnarray}
 where $c_{1}=$ 1.01 fm, $c_{2}= 1. 8$ fm, $V_{1}=$ 61.14 MeV, $V_{2}=-60. 65$ MeV, $W= 1-M$, and $M= 0. 60$.
The singlet-even channel of the original Volkov interaction without the Bartlet ($B$) and Heisenberg ($H$) parameters is known to be considerably strong; thus, $B= H= 0.08$ was introduced to remove the bound state of two neutrons.

For the spin-orbit term, we introduce the G3RS potential \cite{G3RS-1,G3RS-2}, $V_{ls}= V_{0}( e^{-d_{1}r^{2}}-e^{-d_{2}r^{2}}) P(^{3}O) \bm{L} \cdot {\bm{S}}$, where $d_{1}= 5. 0$ fm$^{-2},\ d_{2}= 2. 778$ fm$^{-2}$,
 $V_{0}= 2000$ MeV, and $P(^{3}O)$ is a projection operator onto a triplet odd state.
The operator $\bm{L}$ stands for the relative angular momentum and $\bm{S}$ represents the spin ($\bm{S_{1}}+\bm{S_{2}}$).
All parameters of this interaction were determined from the $\alpha+ n$ and $\alpha+\alpha$ scattering phase shifts \cite{OKA79}.

For the MCC calculation, we calculated the diagonal and transition densities, which are expressed as follows:
\begin{eqnarray}
&&\rho_{Im,I'm'} (\bm{r}) \nonumber \\
&&= \Braket{\Phi^{J^{\pi} M} | \sum_{i=1} \delta (\bm{r}_{i}-\bm{R}_{\rm{c.m.}}-\bm{r}) | \Phi^{(J^{\pi})' M'} } \\
&&= \sqrt{4 \pi} \sum_{\lambda, \mu} (I' m' \lambda \mu | I m) \rho^{(\lambda)}_{I I'} (r) \mathscr{Y}^{*}_{\lambda \mu} (\hat{\bm{r}}), \label{eq:dens}
\end{eqnarray}
where $\mathscr{Y}_{LM}(\hat{\bm{r}}) = i^{L}Y_{LM}(\hat{\bm{r}})$.
Here, $(I' m' \lambda \mu | I m)$ denotes the Clebsch-Gordan coefficient.
$\bm{R}_{\rm{c.m.}}$ is the barycentric coordinate, and $I'$ and $I$ represent the angular momentum of the nucleus for the initial and final states, respectively.
Furthermore, $m$ and $m'$ are the $z$-components of $I$ and $I'$, respectively.
The proton and neutron parts of the densities are separately obtained.
The wave function of the microscopic cluster model, $\Phi^{J^{\pi} M}$, is described as a linear combination of the basis states as in Eq.~(\ref{mcmwf}).
The coefficients for this linear combination are obtained by diagonalizing the Hamiltonian matrix.

\subsection{MCC model}
Next, we move on to the nuclear reaction calculation.
We applied the calculated transition densities of the $^{6}$Li, $^{7}$Li, $^{8}$Li, and $^{9}$Li nuclei to MCC calculations with the complex $G$-matrix interaction MPa~\cite{YAM14,YAM16}.
The MPa interaction has been proven to be successful in heavy-ion scatterings~\cite{FUR16,WWQ17}.
As a detailed calculation procedure for the folding potential has been described in previous research~\cite{FUR13-2}, herein, only the essence of the MCC calculation was briefly introduced.

The diagonal and transition potentials required in the coupled-channel equation were obtained by the folding procedure in the MCC calculation.
The potentials are obtained as the sum of the direct ($U^{(DI)}$) and exchange ($U^{(EX)}$) terms from the microscopic viewpoint as follows:
\begin{equation}
U_{\alpha (ij) \to \beta (kl)} = U^{(DI)}_{\alpha (ij) \to \beta (kl)} + U^{(EX)}_{\alpha (ij) \to \beta (kl)},
\end{equation} 
where $\alpha$ and $\beta$ are the channel numbers and $i, j, k$, and $l$ indicate the states of the projectile and target nuclei.
The direct part of the folding potential is described as
\begin{equation}
U^{(DI)}_{\alpha (ij) \to \beta (kl)} (\bm{R}) = \int{\rho^{(P)}_{i \to k}(\bm{r_p}) \rho^{(T)}_{j \to l}(\bm{r_t}) v_{DI}(\bm{s}, \rho, E/A) d\bm{r_p}d\bm{r_t}}, \label{eq:di}
\end{equation} 
whereas the exchange part is described as
\begin{eqnarray}
U^{(EX)}_{\alpha (ij) \to \beta (kl)} (\bm{R}) &=& \int{\rho^{(P)}_{i \to k}(\bm{r_p}, \bm{r_p}-\bm{s}) \rho^{(T)}_{j \to l}(\bm{r_t}, \bm{r_t}+\bm{s})} \nonumber \\
&& v_{EX}(\bm{s}, \rho, E/A) j_0{\left( \frac{m k(R) s}{M} \right)} d\bm{r_p}d\bm{r_t}, \label{eq:ex}
\end{eqnarray} 
where $\bm{s}=\bm{r_p}+\bm{R}-\bm{r_t}$.
$E/A$ is the incident energy per nucleon. 
Here, $M$ and $m$ are the reduced mass and nucleon mass, respectively.
Note that the Coulomb part of the folding potential was obtained in a manner same as that used for calculating the proton densities of the projectile and target nuclei.
$v_{DI}$ and $v_{EX}$ are the direct and exchange parts of the $N$-$N$ interaction, respectively, for which we adopted MPa~\cite{YAM14,YAM16}.
$j_{0}$ is the spherical Bessel function of rank 0.

In the exchange part of Eq.~(\ref{eq:ex}), $k(R)$ is the local momentum of the nucleus-nucleus relative motion, which is defined as follows:
\begin{equation}
k^2(R)=\frac{2M}{\hbar^2}[E_{\rm{c.m.}}-{\rm{Re}}U^{(\rm{Nucl.})}(R)
-V^{(\rm{Coul.})}(R)], 
\label{eq:kkk}
\end{equation}
where $U^{(\rm{Nucl.})}$ and $V^{(\rm{Coul.})}$ are the nuclear and Coulomb parts of the folded potentials for the elastic channel.
The exchange part of the potential is calculated self-consistently based on the local energy approximation using Eq.~(\ref{eq:kkk}).
In Eq.~(\ref{eq:ex}), the density matrix $\rho (\bm{a}, \bm{b})$ is approximately expanded in the same manner as in previous research~\cite{NEG72}.
\begin{equation}
\rho (\bm{a}, \bm{b}) = \frac{3}{k^{\rm{eff}}_{F}s} j_1 (k^{\rm{eff}}_{F}s) \rho \left( \frac{\bm{a}+\bm{b}}{2} \right) \label{eq:d-mat}
\end{equation}
where $k^{\rm{eff}}_{F}$ is the effective Fermi momentum \cite{CAM78} defined as
\begin{equation}
k^{\rm{eff}}_{F} 
=\left\{ \left(\frac{3\pi^2 \rho}{2} \right)^{2/3}+\frac{5C_{\rm{s}} \left( \nabla \rho \right)^2}{3\rho^2}
+\frac{5\nabla ^2\rho}{36\rho} \right\}^{1/2}, \;\; 
\label{eq:kf}
\end{equation}
and we adopted $C_{\rm{s}} = 1/4$ following a previous study~\cite{KHO01}. 

We employed the so-called frozen-density approximation (FDA)~\cite{FUR09} to evaluate the local density $\rho$ in Eqs.~(\ref{eq:di}) and (\ref{eq:ex}).
In FDA, the density-dependent $N$-$N$ interaction is assumed to feel the local density, which is defined as the sum of the densities of the projectile and target nuclei:
\begin{eqnarray}
\rho &=& \rho^{(P)}+\rho^{(T)}.
\end{eqnarray}
For calculating the potentials, we used the average nucleon densities in the initial and final states for each nucleus~\cite{ITO01,KAT02}:
\begin{eqnarray}
\rho^{(P)} &=& \frac12 \left\{ \rho^{(P)}_{i \to i}+ \rho^{(P)}_{k \to k} \right\} \; , \\
\rho^{(T)} &=& \frac12 \left\{ \rho^{(T)}_{j \to j}+ \rho^{(T)}_{l \to l} \right\} \;.
\end{eqnarray}
The local densities were evaluated at the position of each nucleon for the direct part and at the middle point of the interacting nucleon pair for the exchange part by following the method used in the preceding research~\cite{KHO00,KAT02}. 
FDA has also been widely used in the standard double-folding model calculations~\cite{KHO94, KHO97, KAT02, SAT79, FUR09, FUR14R, FUR16}, it has been proven to be the most appropriate candidate for evaluating the local density in the double-folding model calculations with realistic complex $G$-matrix interactions~\cite{FUR09, FUR16}. 

Although the spin-orbit interaction in the nucleon-nucleon system was considered in all structure calculations, the spin-orbit potential for the Li + $^{12}$C and Li + $^{28}$Si systems was ignored in the present reaction calculation, which has been demonstrated to be negligible for the elastic and inelastic cross sections in previous studies~\cite{SAK86-so, ZAH96}. 
It is worth noting that the analyzing power is also useful for investigating the CC effect, as reported in previous research~\cite{NIS82, NIS84, OHN82,OHN84}.
However, it is difficult to construct the spin-orbit and tensor potentials in our framework.
In this study, the glue-like behavior exhibited by the Li isotope through the CC effect has been only investigated in the context of the elastic cross section.
In addition, the magnetic-multipole transitions (M1 and M3) have not been considered herein.

\vspace{3mm}
\section{Results}
\subsection{Energies, root-mean-square radii, and transition strengths}\label{structure}
\begin{figure}[t]
\begin{center}
\includegraphics[width=6.5cm]{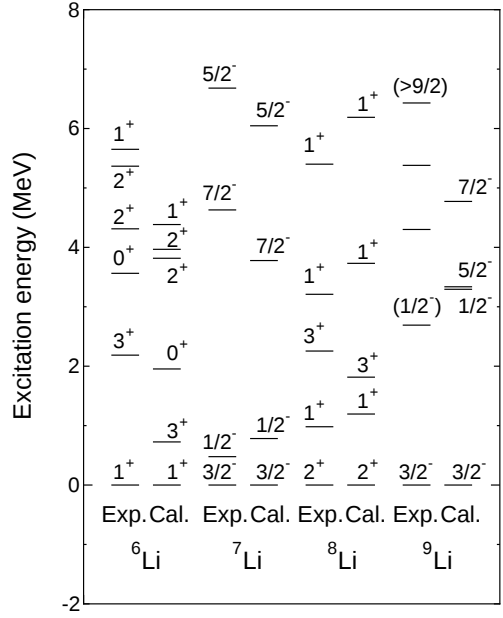}
\end{center} 
\caption{\label{fig:EEE} Excitation energy of the experimental data and calculated results for the Li isotopes.
The experimental data are taken from a previous study~\cite{NNDC}.}
\end{figure}
Firstly, we summarize the results of the structure calculation using the SMCM method.
Figure \ref{fig:EEE} shows low-lying excitation energies for the $^{6}$Li, $^{7}$Li, $^{8}$Li, and $^{9}$Li nuclei compared with the experimental data.
The calculated results agreed well with the experimental data.
The total energies of $^{6}$Li, $^{7}$Li, $^{8}$Li, and $^{9}$Li were $-29.66$, $-37.64$, $-38.35$, and $-41.04$ MeV, respectively.
The excitation energies $E_x$ for the 3$^+_1$, 0$^+_1$, 2$^+_1$, 2$^+_2$, and 1$^+_2$ states of the $^{6}$Li nucleus were 0.73, 1.95, 3.82, 3.97, and 4.38 MeV, respectively.
The calculated energies were slightly lower in comparison with the experimental data; however, the order of each excited state was well reproduced.
The excitation energies for the 1/2$^-_1$, 7/2$^-_1$, and 5/2$^-_1$ states of the $^{7}$Li nucleus were 0.78, 3.77, and 6.05 MeV, respectively,
whereas those for the 1$^+_1$, 3$^+_1$, 1$^+_2$, and 1$^+_3$ states of the $^{8}$Li nucleus were 1.20, 1.82, 3.73, and 6.19 MeV, respectively.
The $3/2^-_1$ and $1/2^-_1$ states ($7/2^-_1$ and $5/2^-_1$ states) for the $^{7}$Li nucleus were basically the spin-orbit partners of $\alpha$ and triton motion with a relative angular momentum of 1 (3), and this structure was well reproduced without assuming a triton cluster a priori.
In the $^{6}$Li and $^{8}$Li nuclei, mixing different spin configurations for the valence neutrons is considerably important.
Additionally, the excitation energies for the 1/2$^-_1$, 5/2$^-_1$, and 7/2$^-_1$ states of the $^{9}$Li nucleus were 3.30, 3.34, and 4.77 MeV, respectively.
Furthermore, the SMCM method predicted the existence of unobserved states below 8 MeV.
The 3$^+_2$, 2$^-_1$, and 1$^-_1$ states for the $^{6}$Li nucleus were found to occur at $E_x$ = 4.73, 5.50, and 6.21 MeV, respectively, and the 1/2$^+_1$ state for the $^{7}$Li nucleus was found to occur at 7.54 MeV.
Furthermore, the 0$^+_1$, 2$^+_2$, 2$^+_3$, and 2$^-_1$ states were predicted for the $^{8}$Li nucleus at $E_x$ = 3.65, 4.32, 5.10, and 5.43 MeV, respectively, and the 3/2$^-_2$ state fir the $^{9}$Li nucleus was predicted at 5.40 MeV.
It should be noted that the resonance condition must be imposed above the threshold.
However, it is difficult to distinguish the resonance and continuum states because the present calculation adopts bound-state approximation.
Herein, we assumed that the calculated state is a resonance state because the obtained states yield standard radii and transition strengths.

\begin{table}[h]
\caption{Calculated root-mean-square (RMS) radii of the charged proton [point proton], point neutron, and point matter for the ground state of the $^{6}$Li, $^{7}$Li, $^{8}$Li, and $^{9}$Li nuclei along with the experimental data.}
\label{tab:01}
  \begin{tabular}{cccc} \hline
                & Charged proton (fm) & Point neutron (fm) & Point matter (fm) \\ \hline  \hline
Exp.~\cite{TAN85} &  &  &  \\ 
$^{6}$Li & 2.54(3) & 2.54(3) & 2.54(3) \\ 
$^{7}$Li & 2.43(3) & 2.54(3) & 2.50(3) \\ 
$^{8}$Li & 2.41(3) & 2.57(3) & 2.51(3) \\ 
$^{9}$Li & 2.30(2) & 2.50(2) & 2.43(3) \\ \hline
Exp.~\cite{SAN06} &  &  &  \\ 
$^{6}$Li & 2.517(30) &  &  \\ 
$^{8}$Li & 2.299(32) &  &  \\ 
$^{9}$Li & 2.217(35) &  &  \\ \hline
Cal. & [Point proton] &  &  \\ 
$^{6}$Li & 2.662 [2.523] & 2.556 & 2.539 \\ 
$^{7}$Li & 2.636 [2.495] & 2.604 & 2.558 \\ 
$^{8}$Li & 2.530 [2.379] & 2.615 & 2.529 \\ 
$^{9}$Li & 2.393 [2.237] & 2.562 & 2.445 \\ \hline
  \end{tabular}
\end{table}
Table~\ref{tab:01} lists the calculated root-mean-square (RMS) radii of the ground states.
The theoretical charged proton radius was obtained by folding the point proton distribution with the proton charge form factor.
The calculated point neutron and point matter radii well reproduced the experimental data.
The current results also reproduced a decrease in the charged proton radius because of the addition of valence neutrons; however, these radii were slightly larger in comparison with the experimental data.
The trend of the decrease in the proton radius exhibited the glue-like behavior of the valence neutrons, which attracted the $\alpha$--$t$ clusters in the Li isotope.

\begin{table}[h]
\caption{Transition strengths from the ground states to the excited states for the $^{6}$Li, $^{7}$Li, $^{8}$Li, and $^{9}$Li nuclei.}
\label{tab:02}
  \begin{tabular}{cc} \hline
Nuclide (initial $\to$ final) & B($IS2$) (fm$^4$) \\ \hline
$^{6}$Li (g.s. $\to$ 3$^+_1$) & 57.13 \\ 
$^{6}$Li (g.s. $\to$ 2$^+_1$) & 54.78 \\ 
$^{6}$Li (g.s. $\to$ 1$^+_2$) & 33.56 \\ \hline
$^{7}$Li (g.s. $\to$ 1/2$^-_1$) & 54.47 \\ 
$^{7}$Li (g.s. $\to$ 7/2$^-_1$) & 96.78 \\ 
$^{7}$Li (g.s. $\to$ 5/2$^-_1$) & 22.06 \\ \hline
$^{8}$Li (g.s. $\to$ 1$^+_1$) & 16.24 \\ 
$^{8}$Li (g.s. $\to$ 3$^+_1$) & 47.94 \\ 
$^{8}$Li (g.s. $\to$ 2$^+_2$) & 7.384 \\ 
$^{8}$Li (g.s. $\to$ 2$^+_3$) & 15.68 \\ \hline
$^{9}$Li (g.s. $\to$ 1/2$^-_1$) & 19.26 \\ 
$^{9}$Li (g.s. $\to$ 5/2$^-_1$) & 19.19 \\ 
$^{9}$Li (g.s. $\to$ 7/2$^-_1$) & 31.55 \\ \hline
  \end{tabular}
\end{table}
Table~\ref{tab:02} lists the calculated strengths (B($IS2$)) of transitions from the ground states to the excited states but for B($IS2$) $>$ 5 fm$^4$.
In the comparison of the $^{7}$Li and $^{9}$Li nuclei, the transition strength between states of the same angular momenta was found to have certainly decreased.
The small transition strength of the $^{9}$Li nucleus was considered to arise due to valence neutrons.
These neutrons exhibited a glue-like behavior, thereby attracting the $\alpha$ and $t$ clusters.
The valence neutrons caused the small radius and consistently resulted in a small transition strength for the $^{9}$Li nucleus.

\subsection{Elastic scattering of Li isotopes}
Next, we introduced the MCC method and calculated the elastic-scattering cross sections for the $^{12}$C and $^{28}$Si targets using the transition densities obtained above.
Note that the imaginary part of the potential obtained using the folding calculation was multiplied by a renormalization factor, $N_W$, as follows:
\begin{equation}
U = V + i N_{\rm{W}} W. 
\end{equation}
Here, $V$ and $W$ represent the real and imaginary parts of the folding model potentials, respectively, and $N_{\rm{W}}$ is the only free parameter in the current MCC calculation.
$N_W$ was used to multiply both the diagonal and off-diagonal potentials.

The transition densities of the target nuclei were calculated in the following manner.
For the $^{12}$C target, the transition density was taken from a previous research~\cite{KAM81}.
For the ground-state density of the $^{28}$Si target, we use the nucleon densities that were deduced from the charge densities \cite{CDENS} extracted from the electron-scattering experiments by unfolding the charge form factor of a proton in the standard way \cite{nuclearSize}.
In addition, we adopted the Bohr-Mottelson-type collective model~\cite{DNR} and a relation based on the $K = 0$ rotational band~\cite{NS} to construct the quadrupole ($\lambda$ = 2) components of the transition density in a manner same as that followed in a previous study~\cite{KHO00}.

Herein, we considered both the projectile (Li isotopes) and target ($^{12}$C and $^{28}$Si) excitations.
The excited 1$_2^+$ (4.38 MeV), 3$_1^+$ (0.73 MeV), 3$_2^+$ (4.73 MeV), 0$_1^+$ (1.95 MeV), 2$_1^+$ (3.82 MeV), 2$_2^+$ (3.97 MeV), 1$_1^-$ (6.21 MeV), and 2$_1^-$ (5.50 MeV) states for $^{6}$Li,
the excited 1/2$_1^-$ (0.78 MeV), 7/2$_1^-$ (3.77 MeV), 5/2$_1^-$ (6.05 MeV), and 1/2$_1^+$ (7.54 MeV) states for $^{7}$Li,
the 2$_2^+$ (4.32 MeV), 2$_3^+$ (5.10 MeV), 1$_1^+$ (1.20 MeV), 1$_2^+$ (3.73 MeV), 1$_3^+$ (6.19 MeV), 3$_1^+$ (1.82 MeV), 0$_1^+$ (3.65 MeV), and 2$_1^-$ (5.43 MeV) states for $^{8}$Li,
and the 3/2$_2^-$ (5.40 MeV), 3/2$_3^-$ (8.18 MeV), 3/2$_4^-$ (10.35 MeV), 1/2$_1^-$ (3.30 MeV), 1/2$_2^-$ (9.03 MeV), 1/2$_3^-$ (13.20 MeV), 5/2$_1^-$ (3.34 MeV), 5/2$_2^-$ (9.08MeV), 5/2$_3^-$ (10.19 MeV), 7/2$_1^-$ (4.77 MeV), and 7/2$_2^-$ (11.60 MeV) states for $^{9}$Li, were considered in our calculation.
The theoretical values of the excitation energy were adopted for the Li isotope.
Even if the values of the excited energies were replaced with experimental ones, the calculated elastic cross sections exhibited negligible changes.
Additionally, the excited 0$_2^+$ (7.65 MeV), 0$_3^+$ (14.04 MeV), 0$_4^+$ (14.88 MeV), 2$_1^+$ (4.44 MeV), 2$_2^+$ (10.30 MeV), 2$_3^+$ (13.25 MeV), 2$_4^+$ (16.54 MeV), and 3$_1^-$ (9.64 MeV) states were considered for the $^{12}$C target.
The excited 2$_1^+$ (1.78 MeV) state was considered for $^{28}$Si.
The calculated results including the excitation effects of both nuclei were referred to as ``full-CC''.
In contrast, the calculated results without excitation effects were referred to as ``1-ch''.
The results for ``$^{12}$C$^*$ only'' and ``$^{28}$Si$^*$ only'' include only the target excitation effect.

\subsubsection{Comparison of calculated results with experimental data}\label{compexp}

\begin{figure}[t]
\begin{center}
\includegraphics[width=6.5cm]{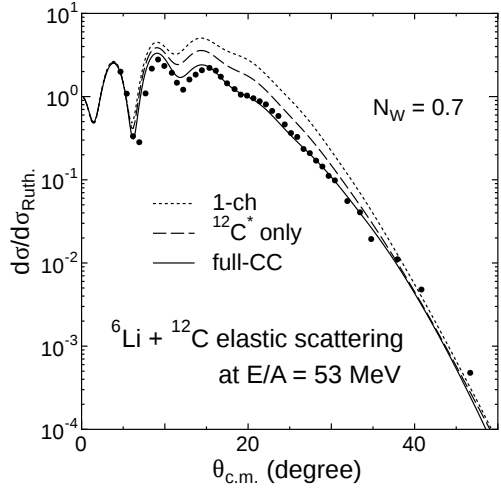}
\end{center} 
\caption{\label{fig:li6c12-53} Elastic cross section for the $^{6}$Li + $^{12}$C system at $E/A$ = 53 MeV.
The experimental data are taken from a previous study~\cite{NAD93}.}
\end{figure}
\begin{figure}[t]
\begin{center}
\includegraphics[width=6.5cm]{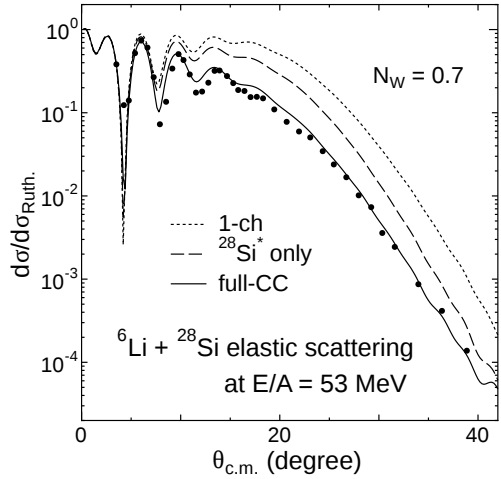}
\end{center} 
\caption{\label{fig:li6si28-53} Same as Fig.~\ref{fig:li6c12-53} but for the $^{28}$Si target.}
\end{figure}
In this section, we compared the calculated results with the experimental data to test our model for light heavy-ion elastic scatterings.
Figures~\ref{fig:li6c12-53} and \ref{fig:li6si28-53} show the elastic cross sections for the $^{6}$Li + $^{12}$C and $^{6}$Li + $^{28}$Si systems, respectively, at $E/A$ = 53 MeV.
The experimental data were well reproduced up to backward angles with $N_W$ = 0.7.
The CC effects for the $^{6}$Li, $^{12}$C, and $^{28}$Si nuclei were clearly apparent on the elastic cross section.
The $^{6}$Li and $^{12}$C nuclei exhibited CC effects comparable those during $^{6}$Li + $^{12}$C scattering.
If the target nucleus is changed to $^{28}$Si, the $^{6}$Li and $^{28}$Si nuclei would exhibit CC effect comparable to those during $^{6}$Li + $^{28}$Si scattering, except for forward angles.
However, according to the visible information, the CC effect during $^{6}$Li + $^{28}$Si scattering appeared to be larger than that during $^{6}$Li + $^{12}$C scattering.

\begin{figure}[t]
\begin{center}
\includegraphics[width=6.5cm]{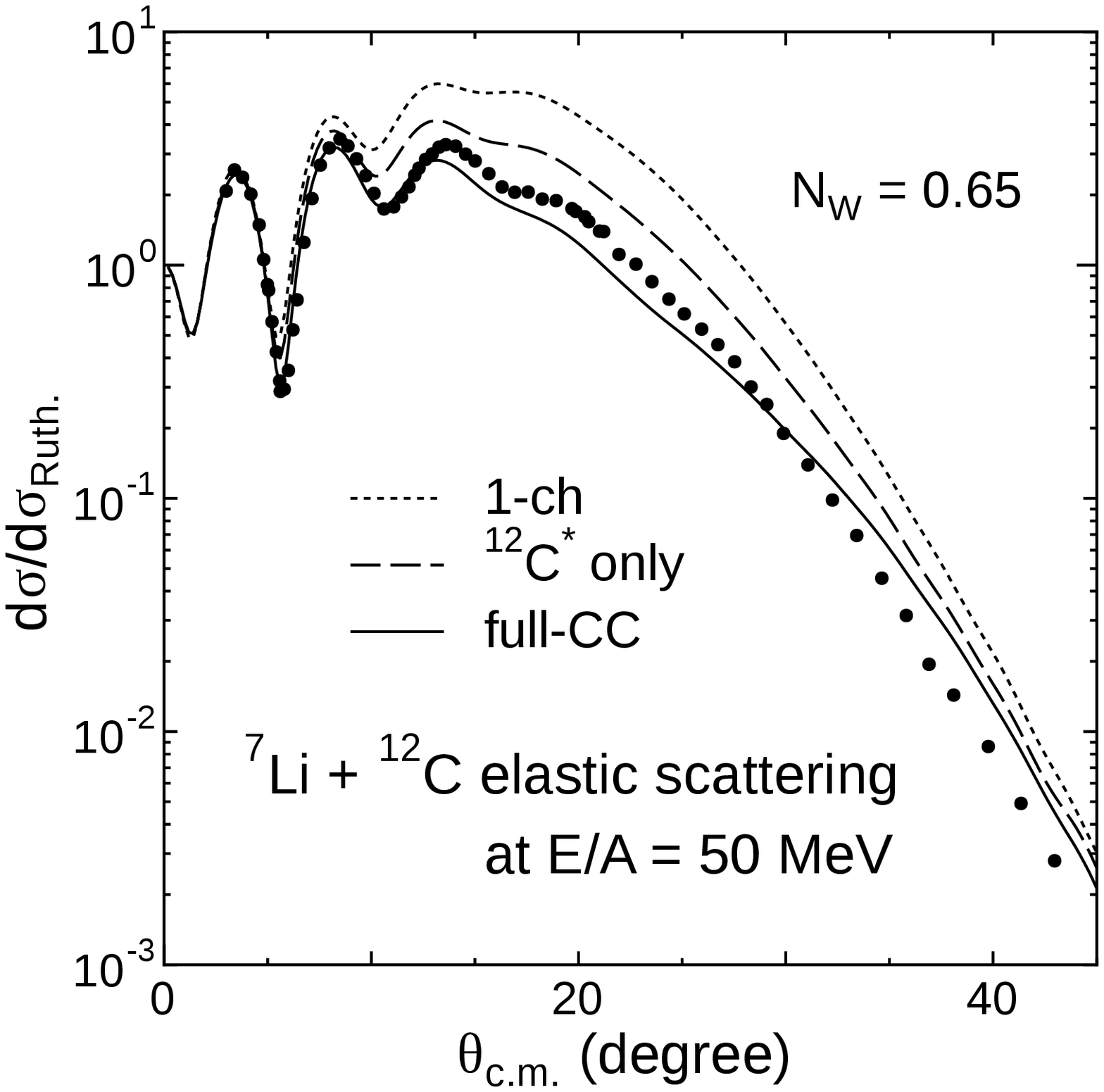}
\end{center} 
\caption{\label{fig:li7c12-50} Elastic cross section for the $^{7}$Li + $^{12}$C system at $E/A$ = 50 MeV.
The experimental data are taken from a previous study~\cite{NAD95}.}
\end{figure}
\begin{figure}[t]
\begin{center}
\includegraphics[width=6.5cm]{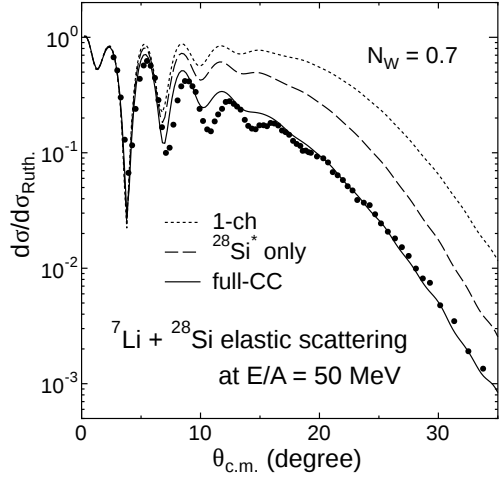}
\end{center} 
\caption{\label{fig:li7si28-50} Same as Fig.~\ref{fig:li7c12-50} but for the $^{28}$Si target.}
\end{figure}
Figures~\ref{fig:li7c12-50} and \ref{fig:li7si28-50} show the elastic cross section for the $^{7}$Li + $^{12}$C and $^{7}$Li + $^{28}$Si systems at $E/A$ = 50 MeV.
The calculated results with $N_W$ = 0.65 and 0.7 were found to reproduce the experimental data, except for the forward angles in the $^{7}$Li + $^{28}$Si system.
A small change in the $N_W$ value in different systems is conceivable in the G-matrix folding model because the complex G-matrix interaction constructed in infinite nuclear matter is applied to the finite nucleus.
Moreover, we confirmed that the other structural model (the orthogonality-condition model~\cite{SAK86}) also required a small change in the $N_W$ value in comparison with the $^{6}$Li and $^{7}$Li elastic scatterings in the present MCC calculation.
The CC effects for the $^{7}$Li, $^{12}$C, and $^{28}$Si nuclei were clearly observed in the elastic cross section.
Again, the CC effect on a system with the $^{28}$Si target appeared to be larger in comparison with that on a system with the $^{12}$C target.

\begin{figure}[t]
\begin{center}
\includegraphics[width=6.5cm]{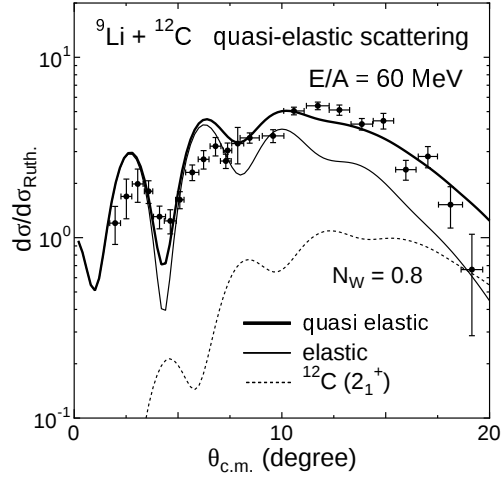}
\end{center} 
\caption{\label{fig:li9c12-60} Quasi-elastic cross section for the $^{9}$Li + $^{12}$C system at $E/A$ = 60 MeV.
The experimental data are taken from a previous study~\cite{PET03}.}
\end{figure}
Next, we changed the nucleus to $^{9}$Li.
The quasi-elastic scattering of the $^{9}$Li + $^{12}$C system at $E/A =$ 60 MeV is shown in Fig.~\ref{fig:li9c12-60}.
The bold curve indicates the incoherent sum of the elastic and inelastic cross sections.
The solid and dotted curves represent the results for the elastic and inelastic cross sections, respectively, of the first excited state (4.44 MeV) of the $^{12}$C nucleus.
The numerical result was slightly different from that reported a previous study~\cite{FUR13-2} because the $N$-$N$ interaction was changed and the numerical procedure was improved.
However, herein, we confirmed that this improvement resulted in only minor changes in comparison with the results reported in previous research~\cite{FUR13-2}.
The experimental data were well reproduced with $N_W$ = 0.8.

\begin{figure}[t]
\begin{center}
\includegraphics[width=6.5cm]{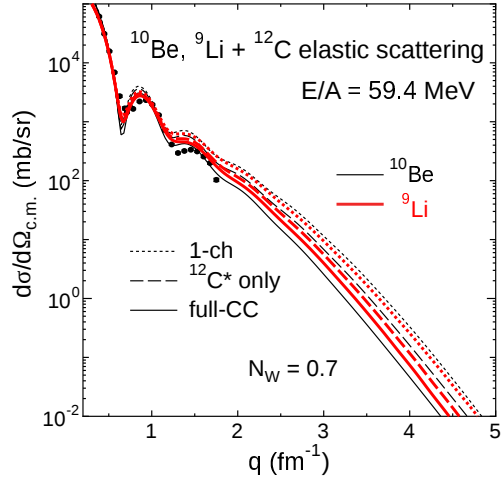}
\end{center} 
\caption{\label{fig:be10c12-59} (Color online) Elastic cross section for the $^{10}$Be + $^{12}$C and $^{9}$Li + $^{12}$C systems at $E/A$ = 59.4 MeV.
The experimental data for the $^{10}$Be + $^{12}$C system are taken from previous research~\cite{EXFOR,COR97}.}
\end{figure}
Furthermore, we compared $^{9}$Li elastic scattering with $^{10}$Be elastic scattering considering the $^{12}$C target.
The experimental data are available for the $^{10}$Be elastic scattering.
The structure of the $^{10}$Be nucleus was described with regard to the $\alpha + \alpha + n + n$ cluster configuration.
The excited energy of the 2$_1^+$ state and the transition strength B$(IS2)$ from the ground state to the 2$_1^+$ state were 3.25 MeV and 234.2 fm$^4$, respectively.
Figure~\ref{fig:be10c12-59} shows the elastic cross section for the $^{10}$Be + $^{12}$C and $^{9}$Li + $^{12}$C systems at $E/A =$ 59.4 MeV.
In this figure, the horizontal axis represent the transfer momentum $q$.
The thin (black) and thick (red) curves represent the results of the $^{10}$Be and $^{9}$Li elastic cross sections, respectively.
The current results were found to well reproduces the data with $N_W$ = 0.7.
The CC effect for the $^{10}$Be nucleus was larger than that for the $^{9}$Li nucleus because the transition strength for the $^{10}$Be nucleus was larger than that for the $^{9}$Li nucleus as reported in a previous study~\cite{FUR17-p1}.
It is worth noting that the numerical results were different from those reported in previous studies~\cite{FUR17-p1, FUR17-p2} because of the improvements in the present calculation.

\subsubsection{Demonstration of CC effect by $^7$Li, $^8$Li, and $^9$Li nuclei}\label{liiso}

As mentioned in the previous subsection, the present calculation well reproduced the experimental scattering data when the renormalization factor was adjusted to around 0.7.
In this subsection, we fixed the renormalization factor to 0.7 and demonstrated the elastic cross section of the Li isotopes.
Note that our conclusions would remain unchanged even if we choose another value near $N_W$ = 0.7 is chosen.
Herein, we reported the elastic cross section of the Li isotopes for the $^{12}$C and $^{28}$Si targets at $E/A$ = 53 MeV.

\begin{figure}[t]
\begin{center}
\includegraphics[width=6.5cm]{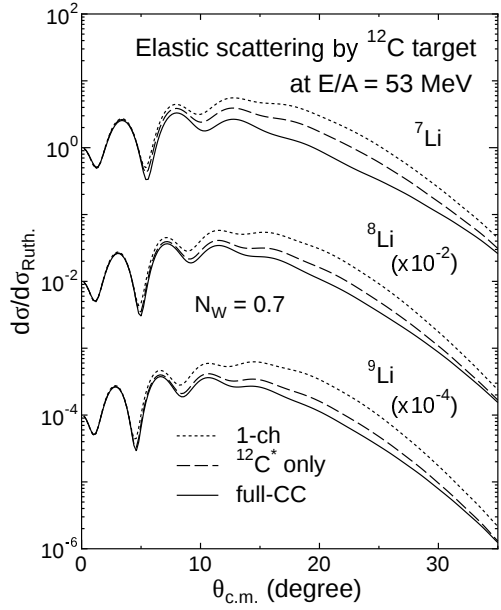}
\end{center} 
\caption{\label{fig:li789c12-53} Elastic scattering of the $^{7}$Li, $^{8}$Li, and $^{9}$Li nuclei for the $^{12}$C target at $E/A$ = 53 MeV.}
\end{figure}
Figure \ref{fig:li789c12-53} shows the $^{7}$Li, $^{8}$Li, and $^{9}$Li elastic scatterings on the $^{12}$C target at $E/A$ = 53 MeV.
We noticed that the CC effects work almost comparably for the projectile and target nuclei during $^{7}$Li + $^{12}$C scattering.
In contrast, a smaller CC effect for the $^{8}$Li and $^{9}$Li nuclei with more neutrons can be observed in Fig.~\ref{fig:li789c12-53}.
Again, this effect indicated that the valence neutrons exhibited an important glue-like behavior, thereby attracting the $\alpha$ and $t$ clusters.
According to the glue-like behavior, the RMS radii of the $^{8}$Li and $^{9}$Li nuclei became smaller.
The shrinkage of the Li isotope resulted in not only a small transition strength but also a small CC effect.

\begin{figure}[t]
\begin{center}
\includegraphics[width=6.5cm]{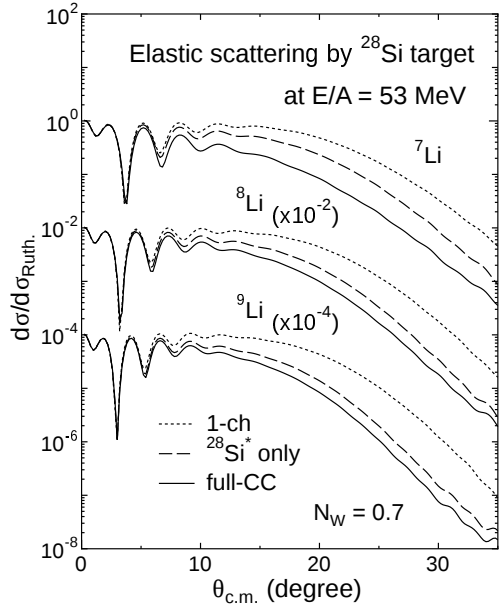}
\end{center} 
\caption{\label{fig:li789si28-53} Same as Fig.~\ref{fig:li789c12-53} but for the $^{28}$Si target.}
\end{figure}
Figure \ref{fig:li789si28-53} shows the $^{7}$Li, $^{8}$Li, and $^{9}$Li elastic scatterings on the $^{28}$Si target at $E/A$ = 53 MeV.
We can observe the large CC effect for the $^{7}$Li nucleus in the $^{7}$Li + $^{28}$Si system.
Furthermore, the small CC effect of the $^{8}$Li and $^{9}$Li nuclei were confirmed.
Again, the results indicate that the valence neutrons exhibited an important glue-like behavior, thereby attracting the $\alpha$ and $t$ clusters in the $^{8}$Li and $^{9}$Li nuclei. 
In the neutron-rich nuclei, we roughly expected that the CC effect will be important since the particle-decay thresholds were always low.
This may be true for the fact that the continuum states for the valence neutrons made significant contribution; however, the valence neutrons also exhibited an aspect of stabilizing the binding of core parts.
Therefore, the CC effect related to core excitation was indeed reduced when we performed a direct comparison of this case with the case involving weakly bound core nuclei.

\section{Summary}
Herein, we combined microscopic structure and reaction frameworks and investigated the channel coupling (CC) effect on the elastic scatterings of the Li isotopes ($A =$ 6--9) for the $^{12}$C and $^{28}$Si targets at $E/A =$ 50--60 MeV.
The wave functions of the Li isotopes were obtained using the stochastic multi-configuration mixing (SMCM) method based on the microscopic cluster model, which yielded reasonable results for the excitation energies and radii in comparison with the experimental data.
The proton radii of the $^{7}$Li, $^{8}$Li, and $^{9}$Li nuclei became smaller with increasing valence neutron additions.
A comparison of the $^{7}$Li and $^{9}$Li nuclei showed that the transition strength [B$(IS2)$] also became smaller because of the addition of valence neutrons.
This indicated that the valence neutrons exhibited an important glue-like behavior, i.e., they were found to bind the $\alpha$ + $t$ clusters in the Li isotope.

With these wave functions, the elastic scatterings of the Li isotopes were obtained in the framework of the microscopic coupled-channel (MCC) method.
The existing experimental data were well reproduced up to backward angles for $A$ = 6--10 elastic scattering by the $^{12}$C and $^{28}$Si targets at $E/A$ = 50--60 MeV.
In addition, the CC effects of the projectile and target nuclei were clearly apparent in the elastic cross sections.
The $^{7}$Li, $^{8}$Li, and $^{9}$Li elastic cross sections were demonstrated for the $^{12}$C and $^{28}$Si targets at $E/A$ = 53 MeV with the inclusion of the CC effect.
The $^{8}$Li and $^{9}$Li nuclei exhibited smaller CC effects in comparison with the $^{7}$Li nucleus, which is thought to be caused by the glue-like behavior exhibited by the valence neutrons; the $\alpha$--$t$ core was stabilized by adding neutrons and the $\alpha$--$t$ distance became smaller.
We realized that the glue-like effect of the valence neutrons on the Li isotope appeared not only in the nuclear radii and the transition strengths but also in the CC effect upon the elastic cross sections.
In the neutron-rich nuclei, we roughly expected that the CC effect will be important since the particle-decay thresholds were always low.
This could possibly be true for the fact that the continuum states for valence neutrons made significant contribution; however, the valence neutrons also tended to stabilize the binding of core parts.
Therefore, the CC effect related the core excitation was indeed reduced when we performed a direct comparison of this case with the case involving weakly bound core nuclei.

\section{Acknowledgment}
This work was supported by Japan Society for the Promotion of Science (JSPS) KAKENHI Grant Numbers JP15K17661, JP15K17662, and JP17K05440.


\end{document}